\documentclass[%
 aip,
 amsmath,amssymb,
 reprint,%
]{revtex4-1}
\usepackage[utf8]{inputenc}
\usepackage{caption}
\usepackage{subcaption}
\usepackage{comment}
\usepackage{bbm}
\usepackage[normalem]{ulem}
\usepackage{amsmath,bm}
\usepackage{amssymb}
\usepackage{graphicx}
\usepackage{booktabs}
\usepackage{xcolor}
\usepackage{xr}
\usepackage{url}

\newcommand{\eb}[1]{{\color{black}{#1}}}

\begin{document}
\preprint{AIP/123-QED}
\title{Inferring the Isotropic-nematic Phase Transition with Generative Machine Learning}
\author{Eric R. Beyerle*\footnote{Corresponding author.}}
\email{eric.beyerle@bio.ku.dk}
\affiliation{Institute for Physical Science and Technology, University of Maryland, College Park, MD 20742, United States}
\author{Pratyush Tiwary*\footnote{Corresponding author.}}
\email{ptiwary@umd.edu}
\affiliation{Department of Chemistry and Biochemistry and Institute for Physical Science and Technology, University of Maryland, College Park, MD 20742, United States}
\affiliation{University of Maryland Institute for Health
Computing, Bethesda, MD 20852, United States}
\date{\today}
\begin{abstract}
    Contemporary work implies generative machine learning models are capable of learning the phase behavior in condensed matter systems such as the Ising model. In this Letter, we utilize a score-based modeling procedure called Thermodynamic Maps to describe the isotropic-nematic phase transition in a melt of $N=343$ calamitic Gay-Berne ellipsoids. When trained on samples generated by molecular dynamics simulation from a single temperature on either side of the phase transition, we demonstrate this generative machine learning approach infers information regarding the critical behavior and estimates effectively the nematic order parameter at sampled temperatures between the two training temperatures. These results demonstrate score-based models' ability to learn the physics of a non-trivial liquid crystalline phase transition driven by anisotropic interactions both entropic and energetic in nature.
\end{abstract}
\maketitle

\emph{Introduction--} Modeling the phase transitions in liquid crystalline systems is important due to their contemporary use in electronic displays, organic electronics, thin-film thermometers, and other useful devices.\cite{deGennes1993,Brown2009,Andrienko2018} In addition to their widespread practical applications, liquid crystals have been the subject of intense experimental, theoretical, and, recently, computational research because they display long-range order along only a single dimension, with a symmetry between that of disordered liquids and completely ordered solids.\cite{Chaikin1995} In general, liquid crystalline phase transitions are seen in systems composed of anisotropic particles or molecules, such as active rods\cite{Doostmohammadi2018}, lipids,\cite{Kulkarni2012, Luzzati1957} and cyanobiphenyls in displays.\cite{Andrienko2018,Palermo2013,Tiberio2009} The physics of liquid crystalline systems are modeled accounting for the constituent particles’ anisotropy in both shape and inter-particle interactions.\cite{Bates1999,Schmid2002,Luckhurt1990,Humpert2016, Hansen2013,Zannoni1994,Ramasubramani2020} An early and successful approach to model not only liquid crystals, but more generally colloidal, anisotropic systems comes via the Gay-Berne (GB) potential,\cite{Gay1981, Everaers2003} which has been used to study the behavior of a variety of colloidal systems for over 40 years.\cite{Jose2004,Andrew1993,Allen1996,Chen2016,Vroege1992} Its ability to model such a wide array of physical systems with computational and analytical efficiency makes the GB model highly useful despite its age.

Effectively an anisotropic Lennard-Jones interaction, the GB potential is parameterized by the relative orientation of the constituent ellipsoids, the ratio of the their long and short axes, and the distance between their centers of mass as follows:
\begin{align}
&U(\widehat{u}_1,\widehat{u}_2,r) =\notag \\
&4\epsilon(\widehat{u}_1,\widehat{u}_2,r)\left[\left(\frac{\sigma(\widehat{u}_1,\widehat{u}_2,\widehat{r})}{r}\right)^{12}-\left(\frac{\sigma(\widehat{u}_1,\widehat{u}_2,\widehat{r})}{r}\right)^6\right],
\label{eq:gb_pot}
\end{align}

where $\widehat{u}_1$ and $\widehat{u}_2$ specify the relative orientation of particles labeled 1 and 2 in a common reference frame; $\widehat{r}$ is the unit vector between the centers of the two ellipsoids; and $r$ is the magnitude of the distance between their centers. The explicit forms of $\epsilon(\widehat{u}_1,\widehat{u}_2,r)$ and $\sigma(\widehat{u}_1,\widehat{u}_2,\widehat{r})$ are given in the original paper,\cite{Gay1981} and interested parties are referred to its implementation in LAMMPS\cite{Brown2009} for additional information regarding the theory and computational implementation of the GB potential. Explicit details regarding implementation of the potential in LAMMPS are given in the SI\cite{si}.

An extensive literature shows this system transitions among liquid crystalline states: isotropic, nematic, smectic, and cholesteric, depending on the thermodynamic conditions.\cite{deMiguel1991,deMiguel1992,Birdi2022,Takahashi2021} Given the simple potential function but rich phase diagram, this model system has great use for the study of condensed matter systems sampling liquid crystalline states. Here, we narrow our focus to the isotropic-nematic phase transition, which is caused by the anisotropy of the GB ellipsoids and is responsible for many optical properties of liquid crystalline displays.

\begin{figure*}[htbp]
     \centering
         \includegraphics[width=0.95\textwidth]{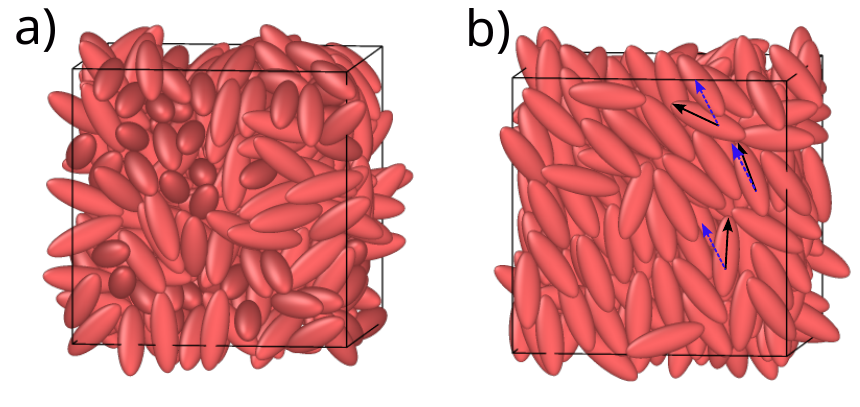}
        \caption{The two phases sampled by the GB system at the phase points studied in this Letter: a) isotropic and b) nematic. Both snapshots are generated using OVITO.\cite{ovito} For the nematic phase in panel b), we show three example orientation unit vectors, $\widehat{u}_i$, as black arrows for three ellipsoids. For these labeled ellipsoids, we also superimpose the approximate director vector, $\widehat{n}$, as a dashed, blue arrow.}
        \label{fig:snapshots}
\end{figure*}

In Figures \ref{fig:snapshots}a,b we show representative configurations of the Gay-Berne melt in the isotropic and nematic phases, respectively. The global difference between the two is the breaking of the O$_3$ rotational symmetry along a protected axial direction called the director,\cite{Chaikin1995} which is superimposed as a blue, dashed arrow on three ellipsoids in Figure \ref{fig:snapshots}b). This orientational symmetry breaking pushes the system from a liquid phase to a liquid crystalline phase and is due to the anisotropic nature of the GB ellipsoids.



We focus on inferring the critical behavior of the isotropic-nematic phase transition given only equilibrium fluctuations in the isotropic and nematic states using an ML approach termed thermodynamic maps (TM).\cite{Herron2023} A score-based approach, TM learns a map between distinct physical states, in this case the isotropic and nematic phases, and infers the physics behind transitioning between these states by learning a fluctuation-coupled backwards diffusion equation.\cite{Herron2023,Anderson1982} A main finding of this Letter is TM's ability to infer successfully the physics of the critical point, where sampling probability is its lowest, based on fluctuations gathered at only two temperatures, both far from the critical point: one temperature where the isotropic phase is stable and another where the nematic phase is stable. 

With the contemporary availability of large training datasets generated through experiment or simulation, statistical techniques relying on machine learning approaches have become popular for identifying and studying phase transitions. These approaches range from shallow, linear ML methods such as PCA for learning the order parameter in condensed matter systems \cite{wetzel_ising, Jadrich2018a,Jadrich2018b,Hu2017}, to more sophisticated, generally non-linear approaches such as convolutional neural networks to identify structure\cite{Bachtis2020,Oh2023} and encoder-based neural networks to learn reaction coordinates for distinguishing phases. \cite{Shen2024, Shen2022, Ccivitciouglu2024,Arnold2024} When using these types of ML approaches, the boundary between using models that are completely interpretable, but too simple to fit the observed data sufficiently well and models that are intractable but yield exceptional fits to the observations is sometimes difficult to straddle. Ideally, we desire a model balancing the bias-variance tradeoff\cite{McGibbon2014}, being physically informed and interpretable yet expressive enough to fit the observed data.

The thermodynamic maps (TM) method we use is a type of generative machine learning architecture called a score-based model\cite{song2020denoising}. These models are nonlinear, data-driven, machine-learned extensions of classical physics-based methods such as targeted free-energy perturbation \cite{Jarzynski2002} and annealed parallel tempering\cite{Neal2001} that map between two probability distributions defined on the same space. Practically, the score-based model is trained to learn the score or the gradient of the target probability distribution by iteratively learning a noising schedule from an estimate of the target distribution to a normal distribution.\cite{Diez2024,wang2022data,sohl2015deep,Bond2021,Bansal2024,song2020denoising,song2020score} The trained model effectively samples from the complicated and multimodal target distribution.

For spin systems, e.g. the Ising model, given samples from a wide range of temperatures, simpler ML models such as PCA and autoencoders can learn the critical behavior of the transition between the para-and ferro-magnetic states.\cite{wetzel_ising,Wang2016,Yevick2022,dAngelo2020}. These calculations require spins from range of temperatures on both sides of the transition as well as at the critical point, reducing their appeal.  However, using TM, a temperature-coupled score-based model, we take one temperature on each side of the critical point and learn the critical behavior of the isotropic-nematic phase transition for a system of GB ellipsoids more effectively than other generative models. This approach guarantees that the ML model balances the bias-variance trade-off by using a complex architecture with sparse inputs to learn the physics at the critical point without overfitting the observed data. For these previous systems, the order parameter demarcating the disordered and ordered states is the global magnetization, a linear sum of the constituent spins. This is also true for previous application of TM to the Ising model.\cite{Herron2023} Here, on the other hand, \eb{calculation of the nematic order parameter requires a non-linear transformation of the ellipsoid orientations, therefore requiring a non-linear modeling approach such as TM.}

\emph{Theory and Methods--} For a GB system comprising $N$ particles, the nematic order parameter $\mathcal{S}$ is defined through an averaging of the local orientation of each GB particle:
\begin{equation}
\mathcal{S}=\Bigg\langle\frac{3\cos^2(\theta)-1}{2}\Bigg\rangle 
:=\frac{1}{N}\sum_i\left(\frac{3\cos^2(\theta_i)-1}{2}\right),
\label{eq:nematic_OP}
\end{equation}
where $\cos(\theta_i)=\widehat{u}_i\cdot \widehat{n}$ is the cosine of the angle between the orientation vector of GB particle $i$, $\widehat{u}_i$, and the director vector $\widehat{n}$. Drawing an analogy to the magnetization order parameter, $M$, of the Ising model, which is defined as $M=\frac{1}{N}\sum_i s_i$, where $s_i\in \{-1,1\}$ is the spin at lattice site $i$, the mapping between the GB and discrete spin model is performed using
\begin{equation}
s_i\longleftrightarrow \cos^2\left(\widehat{u}_i\cdot \widehat{n}\right)=\cos^2\left(\theta_i\right)\label{eq:s_map}
\end{equation}
\begin{equation}
M\longleftrightarrow S:=\frac{3}{2}M_{\text{GB}} + \text{const.};\  M_{\text{GB}}:=\frac{1}{N}\sum_i\cos^2\left(\theta_i\right).\label{eq:M_map}
\end{equation}

The fundamental difference between Ising spins and GB unit orientations is the GB model's invariance to reflections normal to the director axis specified by $\widehat{n}$, which limits the range of $M_{\text{GB}}$ to [0,1] instead of $M\in$[-1,1] in the Ising model. The orientational invariance introduced by taking the dot product changes how the order parameter is calculated from a simple linear average over spins to the average of the square of a cosine function. Analytically, the change of variables introduced in eqs. \ref{eq:s_map} and \ref{eq:M_map} provides a mapping between the models, but this non-linear mapping is non-trivial when learned by an ML model. Crucially, this indicates linear approaches cannot learn this order parameter without at least the kernel trick.\cite{Scholkopf1998} 

The advantage of using a generative model such as a diffusion model over ML classifier methods such as convolutional neural networks or bottleneck-based approaches such as PCA and autoencoders is the ability of the diffusion model to generate novel samples at state points not seen during training. 
To use TM to infer the critical behavior, the model is trained to learn a noising schedule that allows inference of structures generated at a given bath temperature, which is coupled to the fluctuations of the observables. While trained on the one temperature and the associated fluctuations on each side of the critical point, the TM can generalize to unseen temperatures between the two, including the critical point. Thus, TM mixes information from the simulations performed at different temperatures and generates novel samples at a requested and previously unseen temperatures. 

Formally, TM learns a map $\mathcal{M}_{\theta}:\mathbb{R}^{2N}\rightarrow \mathbb{R}^{2N}$ from the $2N$-dimensional space of input configurations and corresponding fluctuations to itself using a neural network with parameters $\theta$. We input the orientation vector $\mathbf{x}_i:=\widehat{u}_i=\left(u^{x}_{i}, u^{y}_{i}, u^{z}_{i}\right)$ for each particle $i$ as a one-dimensional array to a one-dimensional convolutional U-net\cite{Ronneberger2015} architecture. We complement the $N$-dimensional $\mathbf{x}$ with another vector $\bm{\beta}$ of dimensionality $N$ that quantifies the fluctuations of the input features at a given bath temperature. We refer to the vector $\bm{\beta}^{-1}$ as a temperature space, compared to the bath temperature, a scalar. The TM learns the mapping $\mathcal{M}_{\theta}$ that estimates the gradient of the log probability of the sampled data, $\nabla_{\mathbf{x, \bm{\beta}^{-1}}}\log\left(p(\mathbf{x}, \bm{\beta}^{-1})\right)$, also known as the score. This satisfies the following forward and backward diffusion equations (or noising and denoising diffusion models):
\begin{widetext}
\begin{equation}
\label{eq:sde-fwd}
\begin{pmatrix}
    \text{d}\mathbf{x}\\
    \text{d}\bm{\beta}^{-1}
\end{pmatrix}
 = -\frac{1}{2}\sigma(t)
\begin{pmatrix}
    \mathbf{x}\\
    \bm{\beta}^{-1}
\end{pmatrix}
\text{d}t + \sqrt{\sigma(t)}
\begin{pmatrix}
    \sqrt{\bm{\beta}_0^{-1}}\\
    \mathbf{1}
\end{pmatrix}\text{d}\mathbf{w}\quad\text{ and}
\end{equation}

\begin{equation}
\label{eq:sde-bck}
\begin{pmatrix}
    \text{d}\mathbf{x}\\
    \text{d}\bm{\beta}^{-1}
\end{pmatrix} = -\frac{1}{2} \sigma(t) \left[
\begin{pmatrix}
    \mathbf{x}\\
    \bm{\beta}^{-1}
\end{pmatrix} + 
\begin{pmatrix}
    \mathbf{s}_\theta(\mathbf{x},t)\\
    \mathbf{s}_\theta(\bm{\beta}^{-1},t)
\end{pmatrix}\right]\text{d}t + \sqrt{\sigma(t)}
\begin{pmatrix}
    \sqrt{\bm{\beta}_0^{-1}}\\
    \mathbf{1}
\end{pmatrix}\text{d}\mathbf{w},
\end{equation}
\end{widetext}
where $\mathbf{w}\sim \mathcal{N}(0,\mathbf{I})$ and $\sigma(t)$ is the noise schedule of the diffusion model.\cite{ho2020denoising} $\mathbf{s}_\theta(\mathbf{x},t)$ is the score learned by the diffusion model in the input feature space and $\mathbf{s}_\theta(\bm{\beta}^{-1},t)$ is the score learned in the space of the input temperatures. Furthermore, ${\beta}_0$ refers to the prior estimate of the fluctuations at the two bath temperatures in the training data, which are calculated as the sample variance from the input MD data. The additional channel containing information connecting the fluctuations to the temperature allows the TM architecture to infer configurations at novel temperatures unseen during training.\cite{wang2022data}  Details regarding parametrization and training the TM are given in the SI\cite{si} and the associated GitHub repository for this Letter.\cite{github_repo}

The global nematic order parameter $\mathcal{S}$ is estimated from the input features, which are the orientations $\widehat{u}_i$ for each GB particle,  by calculating the largest eigenvalue of the orientation tensor $Q=\frac{1}{N}\sum_i \widehat{u}_i \otimes \widehat{u}_i - \frac{1}{3}I$:
\begin{align}
    \mathcal{S}=\frac{3}{2}\max {\left(\rho(Q)\right)},
    \label{eq:S_eigvals}
\end{align}
where $\rho(Q)$ denotes the eigenvalue spectrum of $Q$\cite{Horn1985} and $I\in \mathbb{Z}^{N\times N}$ is an $N \times N$ identity matrix. The TM model is trained on particle orientations from two temperatures, one below the critical temperature and one above the critical temperature. Then, the trained TM model infers orientations $\widehat{u}_{i,\text{TM}}$ at any given temperature. These inferred orientation are used to calculate $Q$ and $\mathcal{S}$ using eq. \ref{eq:S_eigvals}

We model this weakly first-order phase transition between the isotropic and nematic phases in a GB(3,5,2,1) system\cite{Chen2016,Chakrabarti2006} by performing molecular dynamics (MD) simulations of a melt of $N$ = 343 GB ellipsoids in the NVT ensemble at temperatures ranging from T$^{\star}$=0.2 to T$^{\star}$=2.4, where T$^{\star}$ is the reduced Lennard-Jones (LJ) units of temperature, T$^{\star}$=$k_BT/\epsilon$, with $\epsilon$ the side-side well-depth in the GB potential. At this density and over this span of temperatures, the GB system undergoes a transition from the nematic to isotropic phase at the critical temperature of T${_C}^{\star}\approx1.74$. 

All measured and calculated quantities are reported in reduced LJ units. Phase points are generated along the $\rho^{\star}=0.35$ iso-density line. All simulations are performed using the 10 March 2021 build of LAMMPS.\cite{Thompson2022}

Simulations are initiated by placing the GB ellipsoids in a box of volume $V=34300\sigma$ and ramping the temperature from $T^{\star}=0.1$ to $T^{\star}=3.0$ using a Nos\'{e}-Hoover thermostat.\cite{Shinoda2004} Next, the simulation box is compressed at $T^{\star}=3.0$ using a Nos\'{e}-Hoover barostat to a reduced density of $\rho^{\star}=0.35$. When that density is reached, the system is annealed to the desired T$^{\star}$ before running a production run of 10$^6$ integration steps at the desired state point. The integration timestep is $t^{\star}=0.0015$, giving a total simulation length of 1.5$\times 10^3$. The scaling exponents for GB potential are $\upsilon=1.0$ and $\mu=2.0$; the long axis of the GB ellipsoid is $3.0\sigma$ while the short axis is $1.0\sigma$ for an aspect ratio of 3. Finally, the side-side interactions are set at $1.0\epsilon$ while the end-end interactions are set at $0.2\epsilon$,  making the system studied a GB(3, 5, 2, 1) in the conventional nomenclature\cite{Bates1999}. Details regarding the exact implementation of the potential in LAMMPS are given in the SI.\cite{si}

\begin{figure}
    \centering
    \includegraphics[width=0.45\textwidth]{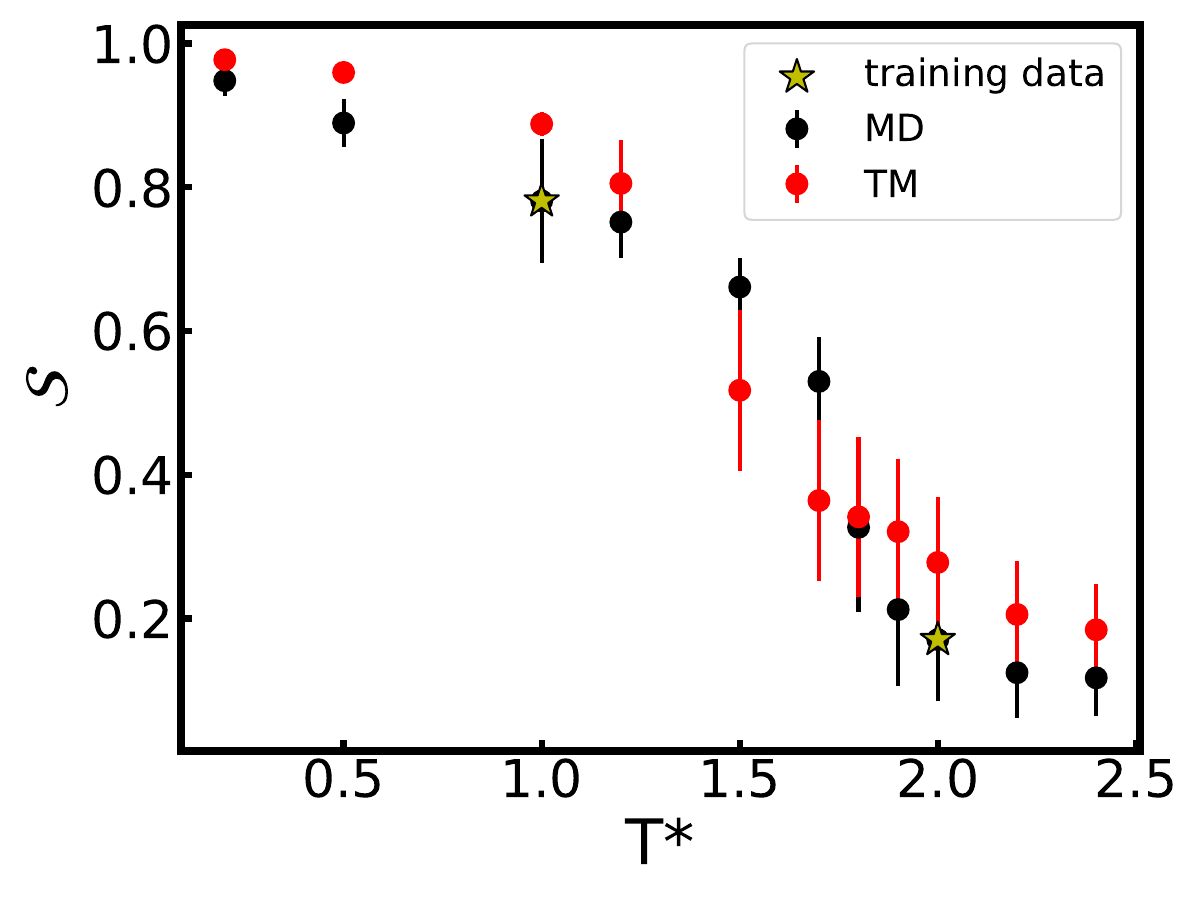}
    \caption{Plot of the nematic order parameter $\mathcal{S}$ calculated from MD (black circles) and the inferred values from trained TM (red circles). The TM results are inferred by training on MD data at the two temperatures highlighted by the yellow stars. Error bars for MD are one standard deviation of $\mathcal{S}$ calculated using 9000 frames from the simulation at the given T$^{\star}$ while the error bars for the TM are taken from 10000 samples inferred at each T$^{\star}$. Error bars for all points correspond to one sample standard deviation.}
    \label{fig:tm_md_S}
\end{figure}

\begin{figure}
    \centering
    \includegraphics[width=0.45\textwidth]{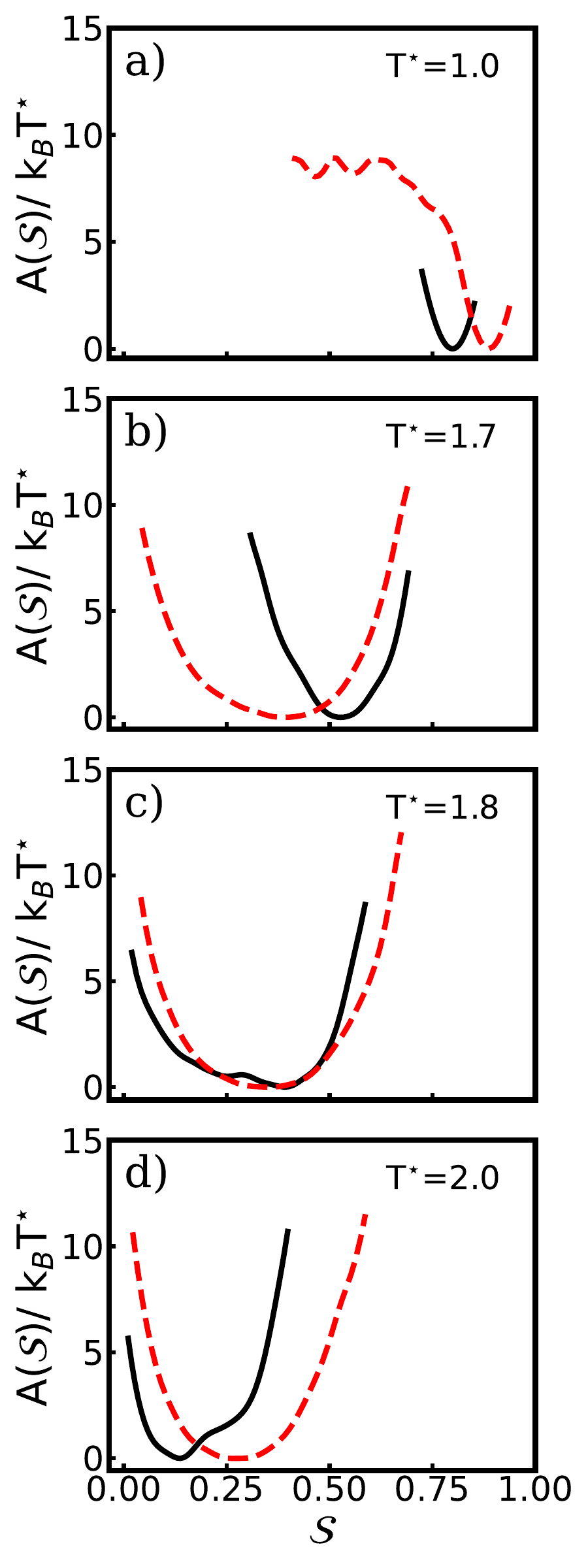}
    \caption{a)-d): Plots of the free-energy along the nematic order parameter $\mathcal{S}$ calculated from the MD (black) and TM (red, dashed) samples at the reduced temperature given in each inset. The temperature increases monotonically going from top to bottom through the subplots. This same plot over a wider sampling of temperatures is given in the SI.\cite{si}}
    \label{fig:P_T}
\end{figure}

\begin{figure}[htbp]
    \centering
        \includegraphics[width=0.45\textwidth]{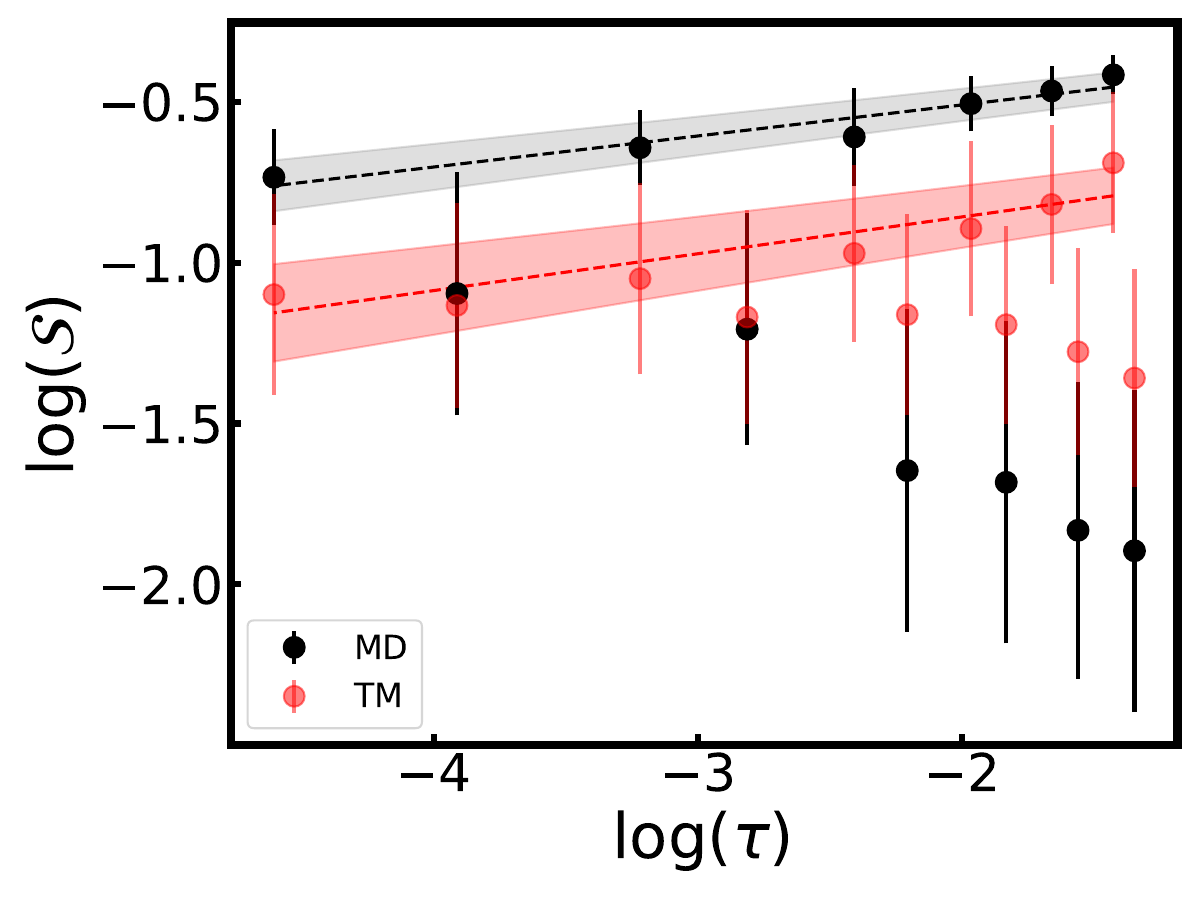}
        \caption{Log-log plot of $\mathcal{S}$ versus $\tau$ with linear fits to the points above the critical temperature shown for both MD (black, dashed line) and TM (red, dashed line) data. Shaded regions show the error in the linear fit, as calculated using the curve fit optimzation function in the Python scipy library.\cite{2020SciPy-NMeth}. Error bars for the MD and TM samples at each temperature are calculated using propagation of error\cite{Taylor1997} from the originally calculated values of $\mathcal{S}$.}
    \label{fig:scaling_laws}
\end{figure}

\emph{Results--} The main result of this Letter is the quantitative and qualitative agreement between the global nematic order parameter $\mathcal{S}$ across temperatures calculated from MD simulations, shown in Figure \ref{fig:tm_md_S}. The TM model making the predictions is trained on only two temperatures far from the critical point. To generate the TM results, 9000 sample configurations at T$^{\star}$=1.0 and T$^{\star}$=2.0 are input to the diffusion model, and 10000 configurations are generated at all temperatures shown in Figure \ref{fig:tm_md_S}. The agreement between the nematic order parameter $\mathcal{S}$ calculated from MD and TM are within error of each other at most temperatures, including temperatures above the highest training temperature T$^{\star}$=2.0. The model struggles to predict the value of $\mathcal{S}$ near the critical temperature of ${\text{T}_{\text{C}}}^{\star}\approx$ 1.74. Even right at the critical temperature the agreement is remarkable when accounting for uncertainty. Furthermore, we are still able to calculate scaling laws in the critical region that agree with those from MD (Figure \ref{fig:scaling_laws} and the agreement between the MD and TM free-energy surfaces near the critical point is relatively good (Figure \ref{fig:P_T}).

As a control, we use a principal component analysis (PCA) \cite{Jolliffe2002} and variational autoencoder (VAE)\cite{kingma2013auto} to learn the nematic order parameter by inputting the projections of each ellipsoid onto the director, $\cos(\theta_i)$, at \emph{all temperatures} into the PCA and VAE architectures. In the SI, we demonstrate that even inputting information at all temperatures, as opposed to only 2 temperatures used to train TM, the linear PCA does not learn any significant physics, always returning a nematic order parameter within uncertainty of a null value. 

The VAE learns an order parameter decaying from the nematic to the isotropic states as temperature increases, but with large error bars, as with PCA. Since the calculation of the order parameter is a nonlinear transformation, it is expected PCA fails to learn a mapping from orientational projection to the order parameter. However, for the VAE, it is a bit surprising that it does not do a better job learning the order parameter even though it utilizes a nonlinear encoder with ReLU activation functions. 

Although the TM cannot match \emph{exactly} the order parameter in the critical region, its performance is vastly superior to the competing PCA and VAE generative models. This result is significantly different from the Ising model, where all three methods effectively learn the phase transition.\cite{Herron2023,wetzel_ising} We emphasize that since the PCA and VAE fail to reproduce $\mathcal{S}(T^{\star})$, even when trained \emph{on all temperatures}, and they completely fail to reproduce critical behavior.

The TM algorithm's ability to match the one-dimensional free-energy surface along the $\mathcal{S}$ order parameter as a function of T$^{\star}$ is shown in Figure \ref{fig:P_T}. For temperatures above and below the critical temperature, the TM always predicts the correct phase: isotropic ($\mathcal{S}\lessapprox 0.5$) above the critical temperature and nematic ($\mathcal{S} > 0.5$) below it. For state points sampled below the critical temperature, the TM approach systematically predicts that the GB melt is more ordered, $\mathcal{S}_{\text{TM}} \ge \mathcal{S}_{\text{MD}}$, compared to the reference MD simulation. This effect could be due to the sensitivity of the use of the dot product in the calculation of the nematic order parameter: practically, there is little physical difference between two GB melts with $\mathcal{S}=0.8$ and $\mathcal{S}=0.9$ when $\mathcal{S}$ is relevant slow variable. 

The free-energy profiles surrounding the transition state are in reasonable agreement with the MD results, with substantial overlapping support between the TM and MD wells, although the TM approach misses some subtleties within the free-energy basin at each temperature. For temperatures above the critical value, the TM method systematically predicts the free-energy minimum is located at a higher value of the nematic order parameter compared to the reference MD simulation, the same phenomenon observed at temperatures below the critical value. The cause of the systematic bias toward higher values of the nematic OP compared to the reference MD simulation is unknown and potential fodder for further study. The only exception to this trend is in the vicinity of the critical point, which is partially elucidated next.

The critical behavior is predicted by fitting scaling laws in the reduced temperature $\tau=\left(\text{T}^{\star} - {\text{T}_C^{\star}}\right)/{\text{T}_C^{\star}}$ for the nematic order parameter $\mathcal{S}$ as follows:
\begin{align}
    S\sim \left|\frac{\text{T}^{\star}-\text{T}^{\star}_\text{C}}{\text{T}^{\star}_\text{C}}\right|^{\mathcal{B}}=\left|\tau\right|^{\mathcal{B}} \label{eq:scaling}
\end{align}
at ranges of T$^{\star}$ exclusively above the critical temperature of T$^{\star}_C\approx$ 1.74. Figure \ref{fig:scaling_laws} compares the scaling laws calculated using the MD and generated TM data. Fitting the branch of the $\mathcal{S}$ versus T$^{\star}$ curve where T$^{\star}>$ T$^{\star}_C$, we report the calculated scaling exponents $\mathcal{B}$ in Table \ref{tab:scaling}. We see $\mathcal{B}$ is in excellent agreement between the TM and MD predictions. 
 
Thus, we conclude the TM method can infer some physics occurring at the critical temperature, as given by the matching predictions for the scaling exponent in Table \ref{tab:scaling}.

\begin{table}[h!]
   \centering
   \caption{Scaling exponents for the nematic order parameter. }
   \label{tab:scaling}
   \begin{tabular}[c]{c c c}
   \hline
   \hline
   &MD& TM \\ \hline
   $\mathcal{B}$ & 0.10$\pm$0.01$^{\text{a}}$& 0.11$\pm$0.03$^{\text{a}}$ \\

    \hline
    \hline
   \end{tabular}
   \\
\textsuperscript{a}{\small Uncertainties are reported from the curve fitting algorithm used from the Python scipy library.\cite{2020SciPy-NMeth}}

\end{table}

Finally, since the TM algorithm, as currently implemented, does not map from the full configuration space of the GB system onto itself, we use a nearest neighbor mapping technique to recover approximate potential energies of inferred TM configurations at each temperature. We assign energies by calculating the minimum distance between a given TM configuration and one in the MD trajectory at the given, sampled temperature. We define the distance between a specified pair of TM and MD configurations as the negative sum of the dot product of the ellipsoid orientations between the two configurations. Formally, he potential energy of TM sample $n$, $U_{\text{TM}}(k)$, is defined as
\begin{equation}
U_{\text{TM}}(k) = U\left(\underset{\mathbf{x}(m)\in\mathbf{X}}{\text{argmin}}\left(-\sum_{i=1}^{N}\left|\widehat{u}_i(m)\cdot{\widehat{u}_i}^{\text{TM}}(k)\right|\right)\right),
    \label{eq:nnmap}
\end{equation}
where $U\left(\underset{\mathbf{x}(m)\in\mathbf{X}}{\text{argmin}}\cdots\right)$ denotes the optimization of the distance between the TM-inferred configurations $\widehat{\mathbf{x}}(k)=\left({\widehat{u}_1}^{\text{TM}}(k),{\widehat{u}_2}^{\text{TM}}(k),\ldots,{\widehat{u}_i}^{\text{TM}}(k),\ldots,{\widehat{u}_N}^{\text{TM}}(k)\right)$ and the MD  configurations $\mathbf{x}(m)=\left({\widehat{u}_1}(m),{\widehat{u}_2}(m),\ldots,{\widehat{u}_i}(m),\ldots,{\widehat{u}_N(m)}\right)$ over all inferred samples $k$ and all MD frames $m$ at a given bath temperature. The TM energy of sample $k$, $U_{\text{TM}}(k)$, is set equal to the MD frame $m$ with which it has the lowest negative sum of dot products between ellipsoid orientational vectors. In the limit where the inputs and outputs of the TM algorithm are the full configuration space of the system, the mapping is exact.

\begin{figure}[htbp]
    \centering
    \begin{subfigure}[b]{0.45\textwidth}
        \caption{}
        \includegraphics[width=\textwidth]{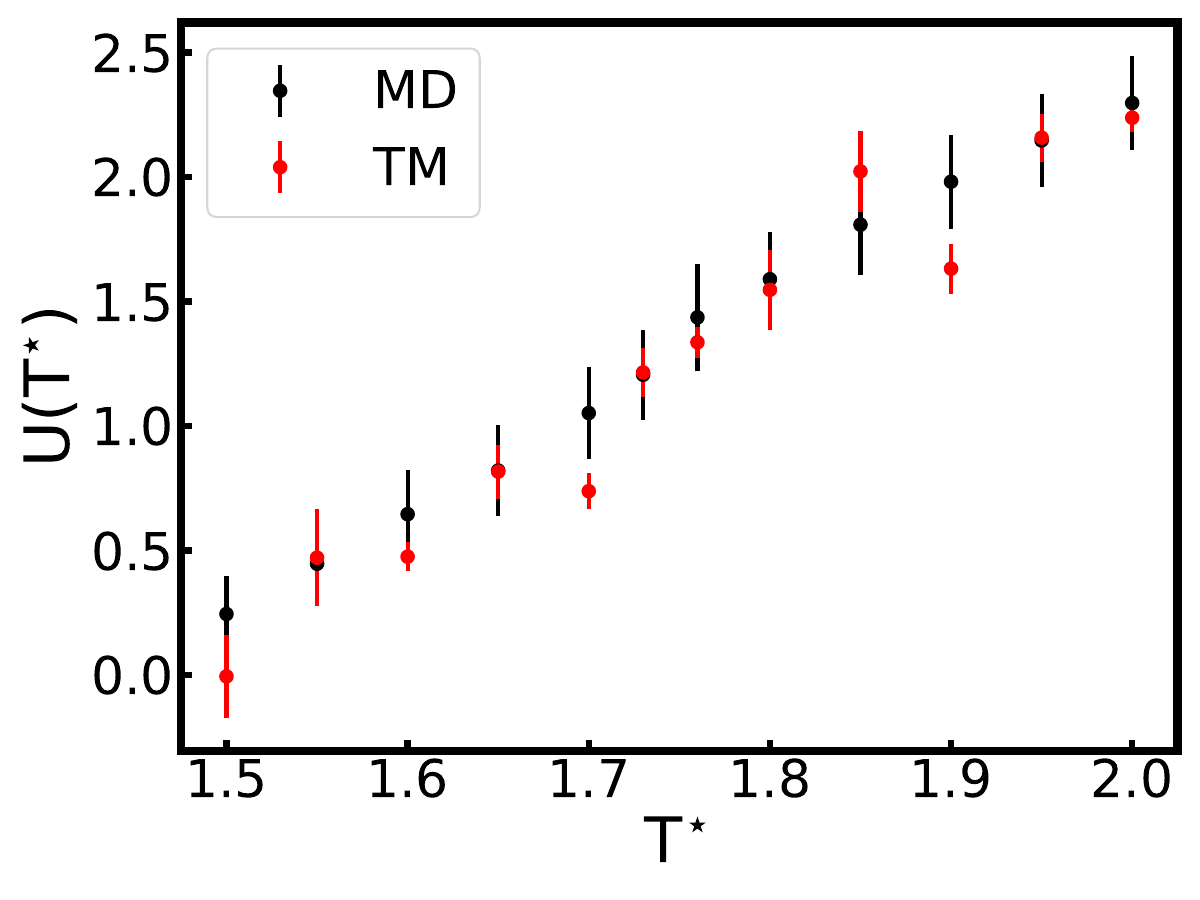}
    \end{subfigure}
    \hfill
    \begin{subfigure}[b]{0.45\textwidth}
        \caption{}
        \includegraphics[width=\textwidth]{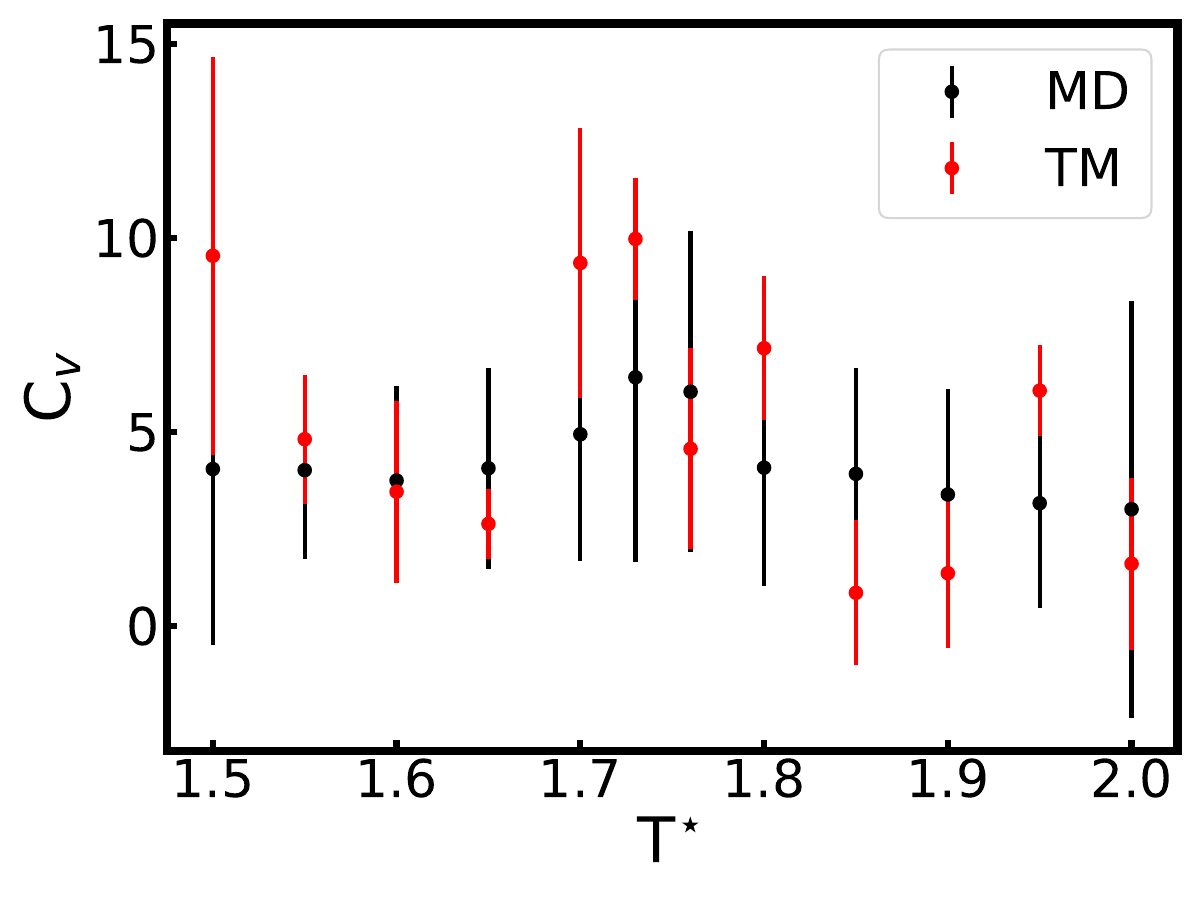}
    \end{subfigure}
    \hfill
    \caption{a) Comparison of the average potential energy at each simulated temperature surrounding the critical region from the reference MD simulations (black) and the inferred energies from the TM approach (red) calculated using the nearest neighbor mapping procedure. Error bars in both cases are given by one standard deviation calculated from the sample variance at each temperature. b) Heat capacity for the reference MD (black) and TM (red) at each simulated temperature surrounding the critical region. The heat capacities for the TM samples are calculated using the mapped energies shown in a). Error bars in both cases are given by one standard deviation calculated from the sample variance at each temperature. }
    \label{fig:U}
\end{figure}

While the energy-matching procedure in Eq. \ref{eq:nnmap} neglects the distance dependence in the GB potential, since the system is largely incompressible at $\rho^{\star}=0.35$, we find this mapping procedure in the reduced space of ellipsoid orientations gives reasonable agreement with the MD energies, even at the critical point, as shown by the plot of energy against temperature in Figure \ref{fig:U}a. 

We use estimate the heat capacity, $C_v\approx\Delta U/\Delta T^{\star}$, from the mapped energies and compare it to the heat capacities measured in the reference MD simulations. For $C_v$ calculated from the MD and mapped TM energies, there is no jump at the critical point, in contrast to the Ising model, where the energy changes sharply at the critical point. \cite{Cowan2005,Herron2023} The gradual change in energy with temperature means there is no peak in $C_v$ at T$^{\star}$ (Figures \ref{fig:U}b and S3), and scaling exponents are within error of zero (Figure S3 and Table S1). 

This behavior of $C_v$ disagrees with data reported in e.g. Ref. \onlinecite{Berardi1993} where there is a clear peak in the heat capacity at the isotropic-nematic transition point. The discrepancy between our qualitative results for both the potential energy and heat capacity curves stem from both the different parameterization of the GB potential and finite-size effects caused by different system sizes used in this study compared to Ref. \onlinecite{Berardi1993}. Indeed, for the Lebwohl-Lasher model, which is the lattice equivalent of the GB system studied here, the peak in the heat capacity is a strong function of system size.\cite{Zannoni2018}

\emph{Conclusions--} In this Letter, we apply a generative, score-matching framework, thermodynamic maps (TM), to infer isotropic-nematic critical behavior in a system of $N$=343 Gay-Berne (GB) ellipsoids. Training on single phase points on either side of the critical temperature, we infer behavior in the neighborhood of the critical point, as indicated by the agreement between the nematic order parameter (Figure \ref{fig:tm_md_S}) and its scaling exponent at criticality (Figure \ref{fig:scaling_laws} and Table \ref{tab:scaling}). 

These results imply the TM method is applicable to non-trivial (i.e. non-Ising) condensed matter systems possessing long-ranged and anisotropic interactions. Such modeling is exceptionally useful in this case since we find that simpler generative modelling procedures such as principal component analysis and the variational autoencoder cannot effectively discriminate the isotropic and nematic phases as successfully as TM even when the ellipsoid orientations are pre-processed to introduce nonlinearity \emph{a priori} (Figure S2). This discrepancy implies the TM's diffusion model, which is effectively a stacked autoencoder\cite{Bishop2023}, is a sufficiently universal approximator of the nematic order parameter from the orientations alone. That is, the TM is able to learn the cosine, squaring, and averaging function to calculate the nematic order parameter from the orientations alone, unlike PCA and the nonlinear VAE. Thus, while a simple model superficially similar to a three-dimensional spin model, the physics of the GB system's liquid crystalline phase transition differs fundamentally from a spin model. Physically, the observation is due to the different symmetries broken for the classical, three-dimensional Heisenberg spin model and the isotropic-nematic phase transition in colloids.\cite{Chaikin1995}

A limitation of this work is the small system size, corresponding to a 7$\times$7$\times$7 lattice. We select this size due to previous use\cite{Chen2016} and computational feasibility. Finite-size effects are obvious when inspecting the asymmetry and rounding in the plot of order parameter versus temperature. \cite{Challa1986, Binder2010} While an issue when comparing the results to theory, our analysis is self-consistent since we compare to MD results. The TM successfully infers some physics at unseen state points despite the system's distance from the infinite size thermodynamic limit.

Since the current implementation of the TM method utilizes the fundamental connection between fluctuations in the relevant degrees of freedom and the system's temperature, we are not able to examine the density-driven nematic-smectic liquid crystalline phase transition.\cite{Polson1997} Updating the TM implementation using a tilted ensemble assumption \cite{Xiao2020, Ohga2024} should allow for a similar treatment of such density-driven phase transitions.

\emph{Acknowledgment--} This work is entirely supported by the US Department of Energy, Office of Science, Basic Energy Sciences, CPIMS Program, under Award DE-SC0021009. We thank UMD HPC’s Zaratan and NSF ACCESS (project CHE180027P) for computational resources. P.T. is currently an investigator at the University of Maryland-Institute for Health Computing, which is supported by funding from Montgomery County, Maryland and The University of Maryland Strategic Partnership: MPowering the State, a formal collaboration between the University of Maryland, College Park and the University of Maryland, Baltimore.

\bibliographystyle{unsrt}
\bibliography{ref}
\end{document}